\documentclass[aps,twocolumn,pra]{revtex4}

\usepackage{graphicx}

\newcommand{\R}{\mathbf{r}}

\newcommand{\UP}{n_{\uparrow}}
\newcommand{\DN}{n_{\downarrow}}

\begin{document}

\title{Assessment of the TCA functional in computational chemistry and solid-state physics}

\author{E. Fabiano}
\affiliation{Istituto Nanoscienze-CNR, Euromediterranean Center for Nanomaterial Modelling and Technology (ECMT), via Arnesano 73100, Lecce}
\affiliation{Center for Biomolecular Nanotechnologies @UNILE, Istituto Italiano di Tecnologia, Via Barsanti, I-73010 Arnesano, Italy}
\author{L. A. Constantin}
\affiliation{Center for Biomolecular Nanotechnologies @UNILE, Istituto Italiano di Tecnologia, Via Barsanti, I-73010 Arnesano, Italy}
\author{A. Terentjevs}
\affiliation{Istituto Nanoscienze-CNR, Euromediterranean Center for Nanomaterial Modelling and Technology (ECMT), via Arnesano 73100, Lecce}
\author{F. Della Sala}
\affiliation{Istituto Nanoscienze-CNR, Euromediterranean Center for Nanomaterial Modelling and Technology (ECMT), via Arnesano 73100, Lecce}
\affiliation{Center for Biomolecular Nanotechnologies @UNILE, Istituto Italiano di Tecnologia, Via Barsanti, I-73010 Arnesano, Italy} 
\author{P. Cortona}
\affiliation{Laboratoire Structures, Propri\'et\'es et Mod\'elisation des Solides, Universit\'e Paris-Saclay, CentraleSup\'elec, CNRS UMR 8580, Grande Voie des Vignes, F-92295 Ch\^atenay-Malabry, France}

\date{\today}

\begin{abstract}
We assess the Tognetti-Cortona-Adamo (TCA) generalized gradient 
approximation correlation functional 
[J. Chem. Phys. 128:034101 (2008)] 
for a variety of electronic systems.   
We find that, even if the TCA functional is not exact 
for the uniform electron gas, it is very accurate
for the jellium surface correlation energies and 
it gives a realistic description of the quantum 
oscillations and surface effects of various jellium clusters,
that are important model systems in computational chemistry and
solid-state physics. 
When the TCA correlation is combined with the 
non-empirical PBEint, Wu-Cohen, and PBEsol$_b$  exchange 
functionals, the resulting exchange-correlation approximations 
provide good performances for a broad palette 
of systems and properties, being reasonably 
accurate for thermochemistry and geometry of molecules, 
transition metal complexes, non-covalent interactions,
equilibrium lattice constants, bulk moduli, and cohesive energies of solids.
\end{abstract}

\maketitle

\section{Introduction}
\label{intro}
Most ab initio electronic structure calculations are 
nowadays performed in the framework of Kohn-Sham
density functional theory (DFT) \cite{hohen,sham}.
The key ingredient of a DFT calculation is the so-called 
exchange-correlation (XC) functional, which
is the only element of the theory that must be approximated.
Thus, the search for efficient XC functionals is a topic of high interest.

There is a great variety of XC approximations. They are usually classified
on the so called Jacob's ladder of DFT \cite{ladder}, which includes
several families of functionals:
local \cite{vwn,pwlda,rc}, semilocal 
\cite{langreth80,pbe,apbe,pbesol,pbeint,zpbeint,q2D,b88,lyp,gaploc,tca,revtca,grac,wuco,tpss,revtpss,js,tpssloc,bloc,blochole},
hybrid \cite{b3lyp,pbe0,pbe0_2,pbe13,guido13,hpbeint,hapbe} 
and fully non-local \cite{kummel_rev,ls2,sosmp2,sos2} functionals.
Each one these families has its own advantages and limitations in terms of accuracy
and computational cost. Hence, no ``best'' or ``universal'' functional
actually exists in practice. Even the simple semilocal functionals display
good utility in many situations.
This is especially true for the generalized gradient approximation (GGA) functionals
that are often the methods of choice when large systems are considered.
For example, in biochemical applications or in solid-state physics,
high-level approaches can hardly be applied,
due to an overwhelming computational cost; instead, quite reliable results
can be obtained at the semilocal level of theory.
Thus, the search for GGAs having broad applicability is
an active research field in DFT 
\cite{hpbeint,zvpbeint,zhao08,m11l,peverati14,peverati12,peverati11,htbs}.

To date, numerous GGAs have been considered in literature. Most of them 
were constructed as whole XC approximations, to benefit of error compensation
between the exchange and correlation contributions. These are indeed quite relevant
at the semilocal level of theory. Nevertheless, some GGA functionals were also
devised as exchange- or correlation-only approximations. In such
cases, the correct association of the exchange functional with the correlation
one is an important issue to allow a practical use of the functional. 

Among the GGA correlation functionals, few variants have gained 
good popularity. In particular, PBE correlation \cite{pbe} 
has been often used as a prototype for non-empirical GGA correlation. 
Similarly famous is the Lee-Yang-Parr (LYP) correlation functional \cite{lyp}, 
obtained as a semilocal approximation
of the Colle-Salvetti \cite{collesal} formula. More recently, correlation
functionals based on the uniform electron gas with a gap have been also 
proposed \cite{gaploc,gap}. Finally, Tognetti, Cortona, and Adamo
built a GGA functional, called TCA \cite{tca}, 
based on the local Ragot-Cortona correlation
\cite{rc} and an average reduced-gradient analysis.

A detailed analysis of the TCA functional is the aim of this paper.
This correlation functional has been the object of only
few studies in the past \cite{tca,thakkar09,peri,agost,radi,labat,pap1,pap2}. 
In particular, no systematic study of the possibility of combining
the TCA correlation functional with one of the existing exchange
functionals has been performed.
Previous studies just showed that using the TCA correlation
together with the PBE exchange \cite{pbe} yields good results for
molecules \cite{tca,peri,agost,radi} but slightly worsens the performances
for solids (with repsect to PBE) \cite{labat}. 
On the other hand, using the TCA correlation
functional together with the PBEsol exchange \cite{pbesol}
some improvement has been achieved for solid-state properties
(especially for cohesive energies), but molecular properties
have been observed to be quite poorly described.

In this paper, we consider these issues and 
we analyze in detail the
behavior of the TCA correlation for different properties
and systems, including jellium spheres and semi-infinite jellium surfaces. 
Furthermore, we assess the possibility
to couple the TCA correlation with some popular
GGA exchange functionals. These tests are conducted
over a fairly large set of molecular and solid-state
properties and systems relevant for semilocal functionals.
Hence, a comprehensive understanding of the
possible performance of the TCA correlation
and its associated XC functionals can be obtained.

\section{Methodology}
To assess the TCA correlation functional we considered its performance
both as a correlation-only functional and in conjuction with different
exchange GGA functionals. 

For the former task we performed a series of tests, 
analyzing the ability of TCA to reproduce the correlation energies
in different systems. In particular, we calculated the correlation energies
of jellium clusters and semi-infinite jellium 
surface energies. For these calculations
we employed accurate LDA Kohn-Sham densities \cite{perdewjell}.
Moreover, we computed the correlation energy of several atoms
and molecules using Hartree-Fock
densities and a cc-pV5Z basis set \cite{cc51,cc52,cc53,cc54}.

To assess the TCA correlation in conjunction with GGA exchange,
we considered a family of PBE-like exchange functionals, which 
includes PBE \cite{pbe}, PBEsol \cite{pbesol}, and
PBEint \cite{pbeint}. We also tested the TCA correlation in association
with several popular GGA exchange functionals,
namely B88 \cite{b88}, OPTX \cite{optx}, Wu-Cohen \cite{wuco},
and a recently introduced variant of the PBEsol exchange (here denoted
PBEsol$_b$) \cite{sll}.
The corresponding XC functionals are denoted
PBE-TCA, SOL-TCA, INT-TCA, B-TCA, O-TCA, WC-TCA, and SOL$_b$-TCA, respectively.

To perform this assessment we employed a large database
covering most of the problems
of interest in quantum chemistry and solid-state
physics that can be described reasonably well
at the semilocal level of theory.
For this reason we do not consider tests that require higher level
treatments, such as dispersion interactions,
dipole moments, and absolute energies.
In detail our test suite can be summarized as follows:
\begin{itemize}
\item \textbf{Main group thermochemistry}: 
atomization energies of small molecules (AE6 \cite{ae6,ae62},
W4 \cite{w4,grimme_test1,grimme_test2},
G2/97 \cite{g2_1,g2_2}), 
barrier heights (BH76 \cite{bh76_1,bh76_2,grimme_test1,grimme_test2}), 
reaction energies (BH76RC \cite{bh76_1,bh76_2,grimme_test1,grimme_test2}, 
OMRE \cite{hpbeint}), and both barrier heights and reaction energies 
(K9 \cite{ae62,k9}).
\item \textbf{Main group geometry}: 
bond lengths of hydrogenic (MGHBL9 \cite{mgbl19})
and non-hydrogenic (MGNHBL11 \cite{mgbl19}) bonds as well as
vibrational frequencies (F38 \cite{f38}) of small organic molecules. 
\item \textbf{Transition metals}:
atomization energies of small transition metal complexes (TM10AE
\cite{3dmetals,zpbeint}) and gold clusters (AUnAE \cite{zpbeint,pbeint_gold},
reaction energies of transition metal complexes (TMRE \cite{hpbeint,3dmetals}),
and bond lengths of transition metal complexes (TMBL \cite{zpbeint,buhl06})
and gold clusters (AuBL6 \cite{hpbeint,pbeint_gold}).
\item \textbf{Non-covalent interactions}: interaction
energies of hydrogen-bond (HB6 \cite{noncov}), 
dipole-dipole (DI6 \cite{noncov}), 
and dihydrogen-bond complexes
(DHB23 \cite{dihydro}).
\item \textbf{Other molecular properties}: difficult cases for DFT (DC9/12 \cite{dc9}), 
small gold-organic interfaces (SI12 \cite{hpbeint}), and 
atomization energies of molecules with non-single-reference character
(W4-MR \cite{w4})
\item \textbf{Solid-state}: Equilibrium lattice constants (LC29),
bulk moduli (BM29), and cohesive energies (CE29) of 29 solids, including
Al, Ca, K, Li, Na, Sr, Ba (simple metals); Ag, Cu, Pd, Rh, V, Pt, Ni (transition metals);
LiCl, LiF, MgO, NaCl, NaF (ionic solids); AlN, BN, BP, C (insulators);
GaAs, GaP, GaN, Si, SiC, Ge (semiconductors). Reference data to construct this
set were taken from Refs. \cite{mattsson08,ss}. 
\end{itemize}

All calculations concerning this test suite, 
except solid-state ones, have been performed
with the TURBOMOLE program package \cite{turbomole} 
using the def2-TZVPP basis set \cite{basis1,basis2} and standard molecular integration
grids (\texttt{gridsize 3} option in TURBOMOLE).
For transition metal atoms, scalar relativistic effective core potentials (ECP)
\cite{ecp1,ecp2} have been employed.
Note that reference data of all tests are corrected for thermal effects, 
thus they are directly comparable with the outcome of the calculations.
In the case of gold clusters, also other relativistic effects 
beyond the ECP treatment are included in the correction \cite{pbeint_gold}.
Fixed reference geometries have been employed
in all molecular calculations.

Solid-state calculations have been performed with the
VASP program \cite{kresse96} using PBE-PAW pseudopotentials.
Note that the use of PAW core potentials ensures good transferability
for multiple functionals \cite{mattsson08}, since the
core-valence interaction is recalculated for each functional.
Indeed, test calculations employing different PAW
potentials have shown that the convergence level of
our results is 
about 1 m\AA{} for lattice constants, 0.5 GPa for bulk moduli, and
0.01 eV for cohesive energies.
All Brillouin zone integrations were performed
on $\Gamma$-centered symmetry-reduced Monkhorst-Pack $k$-point meshes, 
using the tetrahedron method with Bl\"och corrections. 
For all the calculations a $24\times24\times24$ $k$-mesh 
grid was used and the plane-wave cutoff was chosen to be
30\% larger than the maximum cutoff for the pseudopotentials 
of the considered atoms. 
The bulk modulus was obtained using the Murnaghan equation of state. 
The cohesive energy, defined as the energy per atom needed to atomize 
the crystal, was calculated as minus the difference between the total energy
of the crystal at its equilibrium volume and 
the sum of the energies of the constituent atoms as 
obtained from spin-polarized 
symmetry-broken calculations. 
To generate symmetry breaking solutions, atoms were placed in a 
large orthorhombic box with dimensions 13x14x15 \AA$^3$. 
All reference data in the solid-state database are corrected for
zero-point phonon effects.

Calculations for the two example applications, reported
at the end of the paper (see Fig. \ref{fig_final}), were performed
using data and computational setups of Refs. \cite{pbeint,co1}.

To evaluate the overall performance of each functional, 
as well as its accuracy for different classes of problems, each class 
being represented by a set of tests as described above, 
we considered the mean logarithmic relative absolute 
error (LRAE), defined in the following way.
Let $\langle\mathrm{MAE}\rangle_i$ be the average of the 
mean absolute errors (MAE) of the considered functionals 
for the test $i$ (in this work we excluded from the average the best and the
worst functionals). 
The LRAE of a given functional is then defined as 
\begin{equation}\label{eq_lrae}
\mathrm{LRAE}[\mathrm{functional}] = 
\frac{100}{M}\sum_{i=1}^M\log_{10}\left(\frac{\mathrm{MAE}_i[\mathrm{functional}]}
{\langle\mathrm{MAE}\rangle_i}\right)\ ,
\end{equation}
where $M$ is the number of the considered tests. 
This quantity actually summarizes the performances of the functional 
for a heterogeneous
set of data, since it treats adimensional quantities. Note also that
the use of the logarithm makes it independent (up to a rigid shift)
from the $\langle\mathrm{MAE}\rangle_i$ values, in the sense that,  
changing the definition $\langle\mathrm{MAE}\rangle_i$ (e.g including other functionals in the benchmark),
the LRAE of each functional will change, 
but the differences between the LRAEs will be unchanged (thus, the ordering 
will not change).
Hence, more negative values of LRAE indicate that a functional is performing
better than the average of all the considered functionals; 
oppositely, positive values of
the LRAE indicate a performance worse than the average.

\subsection{TCA correlation}
The TCA correlation energy functional \cite{tca} is defined as 
\begin{eqnarray}
\nonumber
E_c^{TCA} & = & \int n\epsilon_c^{TCA}(r_s,\zeta,s)d\R = \\
&& = \int \frac{3}{4\pi r_s^3}\epsilon_c^{RC}(r_s(\R))B(s(\R))C(\zeta(\R))d\R\ ,
\end{eqnarray}
where $r_s=(3/[4\pi n])^{1/3}$, with $n$ being the total
electron density,
\begin{eqnarray}
\nonumber
\epsilon_c^{RC}(r_s) & = & \frac{-0.655868\arctan(4.888270+3.177037r_s)}{r_s}+\\
& &+ \frac{0.897889}{r_s}
\end{eqnarray}
is the local Ragot-Cortona correlation energy per particle \cite{rc},
\begin{equation}
B(s) = \frac{1}{1+\sigma s^\alpha}\ ,
\end{equation}
with $\sigma=1.41$ and $\alpha=2.3$, is an enhancement factor
depending on the reduced gradient 
$s=|\nabla n|/[2(3\pi^2)^{1/3}n^{4/3}]$, and
\begin{equation}
C(\zeta) = \frac{\left[\left(1+\zeta\right)^{2/3} + \left(1-\zeta\right)^{2/3} \right]^3}{2^3}
\end{equation}
is a spin factor (see also Ref. \cite{wang91}), 
with $\zeta=(n_\uparrow - n_\downarrow)/n$ ($n_\uparrow$ and $n_\downarrow$ 
are the spin-up and spin-down electron densities, respectively).

\begin{figure}
\includegraphics[width=\columnwidth]{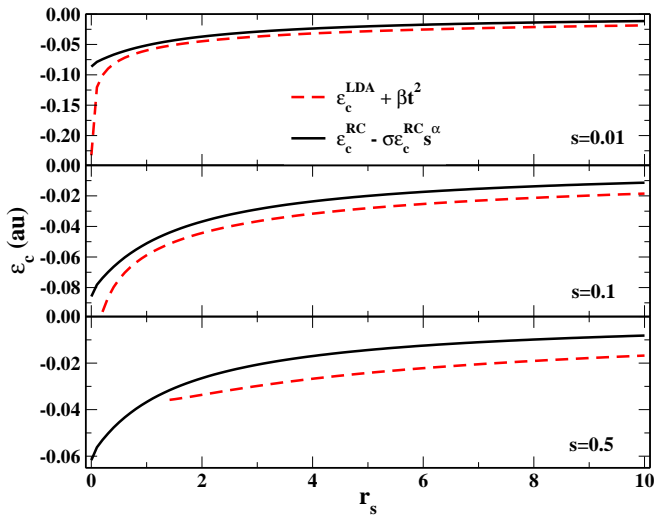}
\caption{\label{fig1} Gradient expansions of the exact correlation and the TCA one for different 
values of the reduced gradient $s=|\nabla n|/[2(3\pi^2)^{1/3}n^{4/3}]$.}
\end{figure}

In the slowly-varying density limit, for the spin-unpolarized case, 
the TCA correlation has the gradient expansion
\begin{equation}\label{e99}
\epsilon_c^{TCA} \approx \epsilon_c^{RC} - \sigma\epsilon_c^{RC}s^\alpha\ .
\end{equation}
This is formally different from the exact one \cite{pbe,revtpss}
\begin{equation}
\epsilon_c \approx \epsilon_c^{LDA} + \beta(r_s)t^2\ ,
\end{equation}
where $\epsilon_c^{LDA}$ is the correlation energy per particle in
the local density approximation (LDA), $\beta$ is the second-order
gradient expansion coefficient, and $t=|\nabla n|/2 k_s\phi  n$ is the reduced gradient for 
correlation \cite{langreth80}, with 
$k_s=(4k_F/\pi)^{1/2}$ being the Thomas-Fermi screening wave vector ($k_F=(3\pi^2n)^{1/3}$), 
and $\phi=\sqrt[3]{C}$ being a spin-scaling factor.
Nevertheless, as shown in Fig. \ref{fig1}, the
slowly-varying behavior of the TCA correlation functional
is similar to the exact one over
a quite large range of values of $r_s$ and $s$.
Thus, the TCA correlation can be expected 
to work fairly well for systems with a slowly varying 
density.
On the other hand, because of the presence of the
exponent $\alpha$ in Eq. (\ref{e99}),
when the TCA correlation is used in conjunction with an exchange 
functional having a gradient expansion of the form
$\epsilon_x\approx\epsilon_x^{LDA}(1+\mu s^2)$, 
i.e. satisfying the second-order gradient expansion
or any of its modifications, 
it is not possible
to enforce exactly the accurate LDA linear response behavior for the
whole resulting XC functional. We recall that this
constraint was instead used in the construction 
of several GGA functionals \cite{pbe,mukappa,hapbe,pbemol}.
\begin{figure}
\includegraphics[width=\columnwidth]{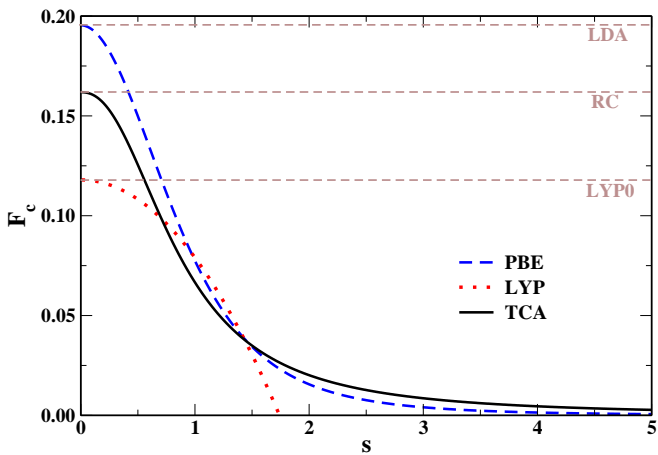}
\caption{Correlation enhancement factor $F_c$ versus the reduced gradient $s$, for $r_s=2$, and for 
$\zeta=0$ (spin-unpolarized case).}
\label{f1}
\end{figure}
In Fig. \ref{f1}, we show the correlation enhancement factor
\begin{equation}
F_c(r_s,s,\zeta)=\epsilon_c(\UP,\DN,\nabla\UP,\nabla\DN)/\epsilon_x^{LDA}(n)\ ,
\end{equation}
versus the reduced gradient $s$, for $r_s=2$, and $\zeta=0$.
At $s=0$, TCA recovers the RC local correlation, PBE \cite{pbe} recovers the
exact LDA correlation, while LYP \cite{lyp} recovers the local correlation present
in this functional (named LYP0).
We recall that both TCA and LYP are based, to some extent,
 on the Colle-Salvetti theory \cite{collesal,lyp}.
Thus, a direct comparison can bring a further insight on the construction
of these functionals.
As shown in Fig. \ref{f1}, TCA decays slower than PBE, but in a similar manner. On the
other hand, the LYP correlation enhancement factor is very different 
especially at large gradients,
where it is strongly negative. This fact implies that $\epsilon_c^{LYP}\geq 0$ 
at large gradients, that is a formally wrong behavior. 
In fact, for systems dominated by large gradients,
such as quasi-two-dimensional systems, the LYP total correlation 
can be positive, failing badly, while TCA performs as PBE.

\section{Results}

\subsection{Assessment of the TCA correlation for atoms and molecules}
The TCA correlation functional has been already assessed for atomic
correlation energies \cite{tca,thakkar09}. 
In Ref. \cite{tca} it was shown that, for the 
H-Ar neutral atoms, the TCA mean absolute error (MAE) on the
correlation energies is 44\% and 17\% smaller than in the case of 
PBE and LYP, respectively.
This improvement was not confirmed for the ions 
Li$^+$-K$^+$: in such a case, TCA is still
better than LYP, but it is slightly worse than PBE \cite{thakkar09}.

To understand better these results, we report in Table \ref{tab_atomc} 
the correlation energies for several
closed- and open-shell atoms and ions as computed from different
DFT functionals.
\begin{table*}
\begin{center}
\caption{\label{tab_atomc} Correlation energy (mHa) divided by the number of electrons ($N_e$) for 
several atoms and ions. Reference data are taken from Refs. \cite{atomref1} and \cite{atomref2}. The 
last lines report the mean error (ME), the mean absolute error (MAE), and the mean absolute relative 
error (MARE).}
\begin{ruledtabular}
\begin{tabular}{lrrrrrrrrrr}
& & \multicolumn{3}{c}{local functionals} & \multicolumn{5}{c}{semilocal functionals} & \\
\cline{3-5}\cline{6-10}
Atom & $N_e$  &  LDA  &  LYP0  &  RC  &  PBE  &  PBEint  &  PBEsol  &  LYP  &  TCA  &  Ref.    \\
\hline
He  &  2  &  -56.2  &  -36.1  &  -47.4  &  -21.0  &  -24.5  &  -26.3  &  -21.9  & -22.4  &  -21.0  \\
Li$^{+}$  &  2  &  -67.3  &  -43.7  &  -56.2  &  -22.4  &  -26.3  &  -28.3  &  -23.8  &  -26.4  &  -21.7  \\
Be$^{2+}$  &  2  &  -75.2  &  -48.2  &  -61.4  &  -23.0  &  -27.2  &  -29.3  &  -24.5  &  -28.6  &  -22.2  \\
Be  &  4  &  -56.0  &  -34.3  &  -45.1  &  -21.4  &  -24.6  &  -26.1  &  -23.6  & -22.2  &  -23.6  \\
B$^{+}$  &  4  &  -63.0  &  -39.0  &  -50.7  &  -23.0  &  -26.5  &  -28.2  &  -26.7  &  -25.1  &  -27.8  \\
C$^{2+}$  &  4  &  -68.5  &  -42.4  &  -54.7  &  -24.0  &  -27.7  &  -29.5  &  -28.6  &  -27.2  &  -35.1  \\
N$^{3+}$  &  4  &  -73.0  &  -45.1  &  -57.8  &  -24.7  &  -28.6  &  -30.5  &  -30.0  &  -28.8  &  -35.1  \\
O$^{4+}$  &  4  &  -76.9  &  -47.3  &  -60.3  &  -25.3  &  -29.2  &  -31.2  &  -30.9  &  -30.0  &  -38.5  \\
Ar$^{8+}$  &  10  &  -96.8  &  -57.1  &  -71.4  &  -41.0  &  -46.1  &  -48.5  & -44.9  &  -46.6  &  -39.9  \\
Ne  &  10  &  -74.3  &  -46.7  &  -59.7  &  -35.1  &  -39.2  &  -41.2  &  -38.4 &  -37.9  &  -39.1  \\
Ar$^{6+}$  &  12  &  -90.2  &  -53.5  &  -67.4  &  -38.3  &  -43.2  &  -45.6  &  -44.8  &  -43.1  &  -41.3  \\
Ar  &  18  &  -79.1  &  -47.9  &  -61.0  &  -39.3  &  -43.5  &  -45.5  &  -41.7  &  -41.5  &  -40.1  \\
Kr  &  36  &  -90.8  &  -52.8  &  -66.6  &  -49.1  &  -53.8  &  -56.0  &  -48.6  &  -50.4  &  -57.4  \\
Zn  &  30  &  -88.5  &  -52.3  &  -66.0  &  -46.9  &  -51.5  &  -53.7  &  -47.7  &  -48.7  &  -56.2  \\
ME  & &  -39.8  &  -10.5  &  -23.3  &  4.6  &  0.5  &  -1.5  &  1.6  &  1.4   &    \\
MAE  & &  39.8  &  11.7  &  23.3  &  5.0  &  4.2  &  4.6  &  3.8  &  4.6    &    \\
MARE  & &  124.81\%  &   41.05\%  &  78.03\%  &  13.02\%  &  12.64\%  &  14.69\%  & 9.92\%  &  13.47\%   &    \\
  &    &    &    &    &    &    &    &    &    &    \\
\multicolumn{11}{c}{Open-shell atoms} \\
Ne$^{7+}$  &  3  &  -80.4  &  -41.8  &  -59.4  &  -19.4  &  -23.2  &  -25.2  & -26.9  &  -27.1  &  -17.0  \\
Be$^+$  &  3  &  -57.6  &  -33.3  &  -46.6  &  -18.1  &  -21.3  &  -23.0  &  -20.4  &  -21.7  &  -15.8  \\
Li  &  3  &  -50.3  &  -29.7  &  -41.2  &  -17.1  &  -20.1  &  -21.6  &  -17.8  & -19.3  &  -15.1  \\
Ar$^{15+}$  &  3  &  -94.9  &  -44.9  &  -64.3  &  -19.7  &  -23.7  &  -25.8  & -29.2  &  -29.1  &  -17.4  \\
C$^{3+}$  &  3  &  -67.7  &  -37.7  &  -53.1  &  -18.9  &  -22.5  &  -24.3  &  -23.7  &  -24.5  &  -16.5  \\
N$^{4+}$  &  3  &  -71.5  &  -39.1  &  -55.2  &  -19.1  &  -22.8  &  -24.7  &  -24.8  &  -25.4  &  -16.7  \\
B$^{2+}$  &  3  &  -63.2  &  -35.8  &  -50.3  &  -18.6  &  -22.0  &  -23.8  &  -22.3  &  -23.3  &  -16.2  \\
O$^{5+}$  &  3  &  -74.9  &  -40.1  &  -56.9  &  -19.2  &  -23.0  &  -24.9  &  -25.6  &  -26.0  &  -16.8  \\
O$^+$  &  7  &  -65.6  &  -38.1  &  -52.8  &  -27.0  &  -30.6  &  -32.4  &  -29.5 &  -30.5  &  -27.7  \\
N  &  7  &  -61.0  &  -35.3  &  -49.4  &  -25.7  &  -29.1  &  -30.8  &  -27.4  &  -28.2  &  -26.9  \\
ME  & &  -50.1  &  -19.0  &  -34.3  &  -1.6  &  -5.2  &  -7.0  &  -6.1  &  -6.9    & \\
MAE  & &  50.1  &  19.0  &  34.3  &  2.0  &  5.2  &  7.0  &  6.1  &  6.9  &      \\
MARE  & &  285.35\%  &  110.24\%  &  196.27\%  &  11.92\%  &  30.48\%  &  40.69\%  & 36.31\%  &  40.53\%    &    \\
  &    &    &    &    &    &    &    &    &    &    \\
\multicolumn{11}{c}{Overall statistics} \\
ME  & &  -44.1  &  -14.0  &  -27.9  &  2.0  &  -1.9  &  -3.8  &  -1.6  &  -2.0    &  \\
MAE  & &  44.1  &  14.8  &  27.9  &  3.8  &  4.6  &  5.6  &  4.8  &  5.6  &       \\
MARE  & &  191.70\%  &  69.87\%  &  127.30\%  &  12.56\%  &  20.07\%  &  25.52\%  &  20.91\%  &  24.75\%    &    \\
\end{tabular}
\end{ruledtabular}
\end{center}
\end{table*}
We see that the local RC functional performs rather well
being almost twice better than the LDA correlation, but still 
worse than LYP0.
On the other hand, the TCA correlation functional
shows an overall performance comparable to the PBEsol
one and slightly worse than LYP and PBE.
However, for closed-shell
systems the TCA correlation performs
similarly to PBEsol and better than PBE.
Moreover, we highlight 
the fact that the TCA functional
generally yields its largest errors
for highly charged ions, i.e.
in high-density limit cases. 
To some extent, this feature is shared 
by all the other GGA 
correlation functionals examined in this paper,
but it appears especially pronounced for TCA and LYP.

To better understand this issue, 
we have performed a reduced-gradient decomposition
of the TCA correlation energy \cite{zupan97}.
This was possible as the variables in the TCA formula are
factorized. We recall that a similar technique was
also used to study kinetic energy functionals
\cite{apbek,l04}.
The TCA correlation energy is thus written as
\begin{equation}\label{sd1}
E_c^{TCA} = \int e_c[n](s)B(s)ds\ ,
\end{equation}
where the $s$-decomposed correlation energy distribution
$e_c$ is defined by
\begin{equation}
e_c[n](s) = \int n(\R)\epsilon_c^{RC}(r_s(\R))C(\zeta(\R))\delta(s(\R)-s)d\R\ .
\end{equation}
This distribution contains all system-dependent information, whereas
the enhancement factor $B$ in Eq. (\ref{sd1}) plays the role of a
universal multiplicative factor.
Using Eq. (\ref{sd1}) we can define the error function
\begin{equation}
\Delta(s) = \int_0^se_c(s')B(s')ds' - E_c^{Ref}\ ,
\end{equation}
with $E_c^{Ref}$ the reference correlation energy.
Thus, the error on the correlation energy is
\begin{equation}
\Delta E_c = \Delta(\infty)\ .
\end{equation}

\begin{figure}
\includegraphics[width=\columnwidth]{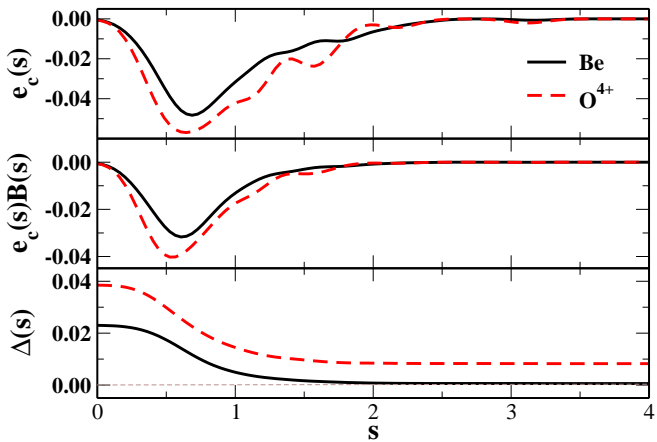}
\caption{\label{sdec_fig} Reduced-gradient decomposition of the TCA correlation energy and its error for the Be 
atom and the O$^{4+}$ ion.}
\end{figure}

The reduced-gradient decomposition of the TCA correlation
energy and the related error are reported
in Fig. \ref{sdec_fig} for the relevant
cases of the Be atom and the O$^{4+}$ ion.
Both systems have 4 electrons but
in the former case TCA performs quite well,
while it gives a relatively large error for O$^{4+}$.
We see that both systems display a similar overall shape for the
$s$-decomposed correlation energy distribution, but
the the curve is more structured for $s>1$ for the O$^{4+}$ ion 
(top panel).
However, the enhancement factor $B$ decays in a rather
fast way ($B(1)\approx0.4$ and $B(2)\approx0.2$) thus
most features at $s>1$ are largely dumped
and the two curves for $e_cB$ are very similar
(middle panel).
As a consequence the error functions $\Delta(s)$
display similar shapes and slopes for both Be and O$^{4+}$
(bottom panel). The only notable difference between the 
two curves is therefore the reference correlation energy,
which is larger for O$^{4+}$ than for Be, causing a shift of the
two lines.
As a consequence for $s\rightarrow\infty$ the
Be curve tends to zero, while the O$^{4+}$ does not,
differing from it approximately by the difference 
between the O$^{4+}$ and Be reference correlation energies.

This analysis suggests that the limitations
of the TCA correlation functional for highly-charged
ions depend mainly on a too fast decay of the enhancement factor, whereas
the RC local functional plays a minor role.
Anyway, we remark that 
Be isoelectronic series is a difficult example of 
strong correlation in the high-density limit \cite{sll}, 
and all semilocal correlation functionals perform poorly
in this case.  

As an additional test for the TCA correlation functional
we report in Table \ref{tab_molc} the correlation energies
for several closed-shell molecules as computed with different
semilocal functionals (local functionals are not reported here
because they perform poorly).
\begin{table}
\begin{center}
\caption{\label{tab_molc} Correlation energy (mHa) for several closed-shell molecules. Reference data are taken 
from Ref. \cite{molc}. The last lines report the mean error (ME), the mean absolute error (MAE), and the mean 
absolute relative error (MARE).}
\begin{ruledtabular}
\begin{tabular}{lrrrrrr}
Molecule  &  PBE  &  PBEint  &  PBEsol  &  LYP  &  TCA  &  Ref  \\ 
\hline
H$_2$  &  -44  &  -49  &  -52  &  -38  &  -41  &  -41  \\ 
LiH  &  -85  &  -98  &  -105  &  -89  &  -88  &  -83  \\ 
Li$_2$  &  -125  &  -144  &  -153  &  -133  &  -130  &  -124  \\ 
CH$_2$  &  -225  &  -251  &  -264  &  -232  &  -228  &  -239  \\ 
CH$_4$  &  -301  &  -332  &  -348  &  -294  &  -298  &  -299  \\ 
H$_2$O  &  -329  &  -364  &  -382  &  -341  &  -339  &  -371  \\ 
FH  &  -339  &  -378  &  -396  &  -363  &  -359  &  -389  \\ 
HCN  &  -442  &  -490  &  -514  &  -463  &  -455  &  -515  \\ 
CO  &  -451  &  -502  &  -527  &  -484  &  -472  &  -535  \\ 
CO$_2$  &  -744  &  -824  &  -863  &  -790  &  -777  &  -535  \\ 
N$_2$  &  -454  &  -504  &  -528  &  -483  &  -472  &  -549  \\ 
SiH$_2$  &  -555  &  -619  &  -649  &  -597  &  -583  &  -567  \\ 
SiH$_4$  &  -615  &  -685  &  -718  &  -648  &  -639  &  -606  \\ 
PH$_3$  &  -643  &  -713  &  -747  &  -675  &  -668  &  -652  \\ 
ClH  &  -688  &  -762  &  -797  &  -727  &  -723  &  -707  \\ 
  &    &    &    &    &    &    \\ 
ME  &  12  &  -34  &  -55  &  -10  &  -4  &   \\ 
MAE  &  42  &  50  &  59  &  42  &  41  &   \\ 
MARE  &  8.9\%  &  12.5\%  &  15.7\%  &  9.2\%  &  8.6\%  &   \\ 
\end{tabular}
\end{ruledtabular}
\end{center}
\end{table}
In this case we find that TCA performs very well
yielding the smallest MAE and MARE, although these are very close to
the PBE and LYP ones.

The results of this section indicate that, for atomic and molecular systems,
the TCA correlation functional is competitive with the popular 
PBE and LYP ones. This is not highly surprising, considering that 
the TCA functional have been constructed using atomic systems
as references. Nevertheless, the present assessment provides a more quantitative
indication on this behavior, confirming that 
the use of TCA for computational chemistry problems
is a viable option.

\subsection{Assessment of TCA correlation for jellium model systems}
Having assessed the TCA correlation functional for atoms and molecules 
we consider here the opposite limit and test the performance of the 
functional for the jellium model. Thus, we report in 
Table \ref{tab2b} the average relative errors of the 
correlation energy with respect to diffusion Monte Carlo (DMC) benchmark values
\cite{perdewjell} ($(E_c-E_c^{DMC})/E_c^{DMC}$) for
magic jellium spheres with $N=2$, 8, 18, 20, 34, 40, 58, 92 and 106 and several 
bulk parameter values (our results agree within 1 mHa with the ones of 
Ref. \cite{perdewjell}).
\begin{table*}
\begin{center}
\caption{\label{tab2b} Average relative errors of the correlation energy ($(E_c-E_c^{DMC})/E_c^{DMC}$) of magic 
jellium spheres (with $N=2$, 8, 18, 20, 34, 40, 58, 92 and 106), for 
several bulk parameters. The fixed-node corrected DMC data are taken from Ref. \cite{perdewjell}.}
\begin{ruledtabular}
\begin{tabular}{lrrrcrrrrr}
 & \multicolumn{3}{c}{local functionals} & $\;\;$ & \multicolumn{5}{c}{semilocal functionals} \\
\cline{2-4}\cline{6-10}
$r_s$ & LDA & RC & LYP0 & & PBE & PBEsol & PBEint & TCA & LYP  \\
\hline
1    & 40.5 & 18.5 & -11.3 & & 6.8 & 13.3 & 11.3 & -6.5 & -22.0 \\
2    & 34.2 & 8.1  & -22.5 & & 7.9 & 13.0 & 11.4 & -12.5 & -29.6 \\
3.25 & 29.9 & -1.7 & -32.2 & & 7.4 & 11.9 & 10.5 & -19.5 & -36.4 \\
4    & 28.1 & -6.3 & -36.3 & & 7.1 &  11.2 & 9.9 & -22.9 & -39.5 \\
5.26 & 26.8 & -12.6& -41.9 & & 7.8 & 11.5 & 10.3 & -27.6 & -43.6 \\
Average & 31.9 & 1.2 & -28.8 & & 7.4 & 12.2 & 10.7 & -17.8 & -34.2 \\
\end{tabular}
\end{ruledtabular}
\end{center}
\end{table*}

We see that the RC results are remarkably accurate, being smaller than the 
exact ones for $r_s=1$ and $r_s=2$ and larger than the exact ones for $r_s\geq 3$. 
The signed average error is 1.2\%. Much larger errors are found for the other
local functionals, including LDA, which is exact for bulk jellium but
rather poor for other density regimes.
Concerning the GGA functionals, we note that all the correlation functionals that 
recover the exact LDA (PBE, PBEint, PBEsol)
give correlation energies smaller than the DMC results, whereas TCA and LYP
results are all larger than the DMC ones. In the case of TCA this behavior is
due to the fact that RC results are already very accurate and $E_c^{TCA}\geq E_c^{RC}$.
In analogy with the case of highly charged ions, this indicates a too
fast decay of the TCA enhancement factor in rapidly-varying regions.
Consequently, TCA results are not very good for jellium clusters, 
although they are also not far from PBEsol ones. 

\begin{figure}
\includegraphics[width=\columnwidth]{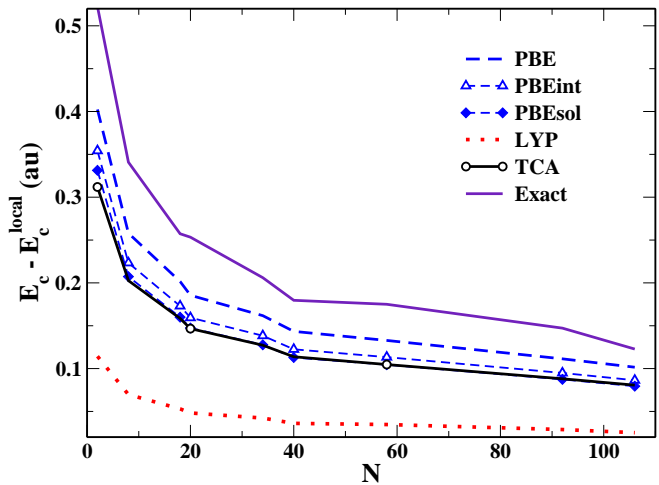}
\caption{ $E_c-E_c^{local}$ for magic jellium spheres with $r_s=2$, and $N$= 2, 8, 18, 20, 34, 40, 58, 92 and 106.} 
\label{f3b}
\end{figure}

Total correlation energies, however, are not the only important feature to
consider in jellium clusters. 
Instead, energy differences
are often more important quantities to consider. In particular, 
we can mention the description of the non-local effects,
which are dominated by quantum oscillations and surface effects and play
a fundamental role in many cases. To investigate this aspect we report 
in Fig. \ref{f3b} the quantity $E_c-E_c^{local}$ 
for magic jellium spheres with 
$r_s=2$, and $N=2$, 8, 18, 20, 34, 40, 58, 92 and 106.
We see that, in this case, all GGA functionals, except LYP, perform 
quite similarly. The TCA results are almost the same as the PBEsol ones
(the two lines are superimposed in the plot) and are very close to 
those given by PBEint. Slightly better results are given by PBE.

To complete our assessment based on the jellium model, we report in Table \ref{tab1}
semi-infinite jellium surfaces correlation energies 
as computed with several functionals.
\begin{table*}
\begin{center}
\caption{\label{tab1} Semi-infinite jellium surface correlation energies (erg/cm$^2$) for different values of 
the bulk parameter $r_s$ as computed with different functionals. Diffusion Monte Carlo (DMC) results \cite{js} 
are also given as reference. The results within the DMC error uncertainty are highlighted in bold style.}
\begin{ruledtabular}
\begin{tabular}{lcrrrcrrrrrcr}
      &         & \multicolumn{3}{c}{local functionals} & & \multicolumn{5}{c}{semilocal functionals} & & \\
\cline{3-5}\cline{7-11}
$r_s$ & $\quad$ & LDA & RC & LYP0 & $\quad$ & PBE & PBEint & PBEsol & TCA & LYP &  $\quad$ & DMC \\
\hline
2 & & 318 & 325 & 267 & & 827 & {\bf 745} & 708 & {\bf 734} & 388 & & 768$\pm$50 \\
3 & &  95 &  96 & 74 & & 275 & {\bf 234} & {\bf 246} & {\bf 243} & 109 & & 242$\pm$10 \\
4 & &  39 &  38 & 27 & & 124 & {\bf 111} & {\bf 105} & {\bf 108} & 40 & & 104$\pm$8 \\
6 & &  10 &   9 & 6 & &  40 &  35 & {\bf 33} &  {\bf 33} & 8 & &  33$\pm\cdots$\\
\end{tabular}
\end{ruledtabular}
\end{center}
\end{table*}
Inspection of the data shows that all 
the local functionals are rather inaccurate for this problem.
Nevertheless, it is interesting to remark that, even if both 
RC and LYP0 are based on the Colle-Salvetti theory,
the RC functional gives results very similar to LDA, while LYP0 is 
even more inaccurate. 

Definitely improved results are given
by all the GGA functionals. Among these, the TCA correlation functional
yields the best performance,
slightly outperforming even PBEsol that was fitted to this property. 
On the other hand, PBE gives larger 
results, while LYP strongly underestimates these jellium surface correlation 
energies. 

The results of this section indicate that the TCA correlation
is not very accurate for total energies of jellium models,
probably because it does not recover the exact LDA limit.
Nevertheless, it performs surprisingly well for energy differences,
such as in the case of semi-infinite jellium surface correlation energies, being
at least competitive with more popular functionals for 
solid-state such as PBEsol.
These results suggest that the TCA correlation, in contrast to the 
LYP one, can also be used in calculations where the 
slowly-varying density regime is relevant.

\subsection{Exchange-correlation functionals}
\label{sec_iv}
The results of the previous subsections indicate that the TCA correlation
functional can be a valid tool both for computational chemistry
and solid-state calculations. 
However, to obtain a practical tool for such applications it is
necessary to couple the TCA correlation with an appropriate exchange functional.
Previous studies \cite{tca,peri,agost,radi,labat} showed that
TCA works well with PBE exchange for molecular problems but not
for solid-state; on the opposite, it performs well
in solid-state together with PBEsol exchange but, in this case, it
does not yield very good results for chemical tests.
Some sort of compromise seems therefore necessary to ensure 
good accuracy for all problems and obtain an XC functional of broad applicability.
To investigate this issue we consider here the performance of
the TCA correlation functional associated with a family of exchange
functionals whose enhancement factor (defined as $F_x=\epsilon_x/\epsilon_x^{LDA}$
with $\epsilon_x$ the exchange energy per particle) has the general form
\begin{equation}\label{aaa}
F_x(s,\alpha) = 1 + \kappa - \frac{\kappa}{1+\frac{\mu(s,\alpha)}{\kappa}s^2}
\end{equation}
where
\begin{equation}
\mu(s,\alpha) = \mu^{GE2} + \left(\mu^{PBE} - \mu^{GE2}\right)\frac{\alpha s^2}{1+\alpha s^2}\ ,
\end{equation}
with $\mu^{GE2}=10/81$, $\mu^{PBE}=0.21951$, and $\alpha$ being a parameter going from
0 to $+\infty$ which controls the bias of the functional towards the slowly-
or the rapidly-varying density regime.
This family includes as limiting cases the PBEsol exchange ($\alpha=0$) and
the PBE exchange ($\alpha=+\infty$). For $\alpha=0.197$ it recovers instead the PBEint
exchange functional \cite{pbeint}.

\begin{figure}
\includegraphics[width=\columnwidth]{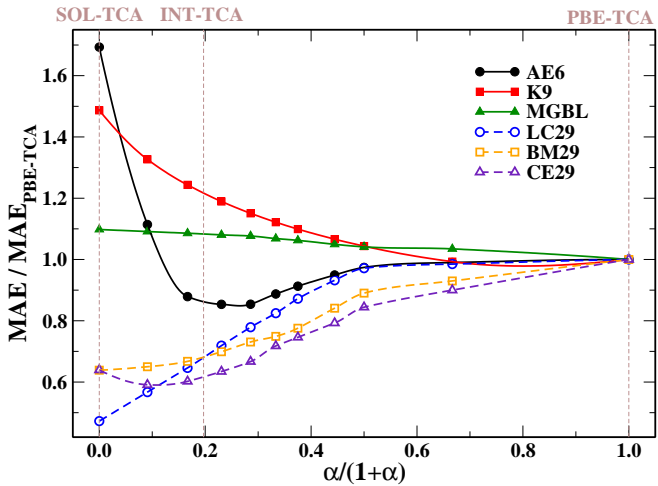}
\caption{\label{scan_fig} Mean absolute errors (MAE) relative to the PBE-TCA one, computed with the XC 
functional using the exchange of Eq. (\ref{aaa}) with different values of $\alpha$ and the TCA correlation, for 
several properties: atomization energies of small molecules (AE6) \cite{ae6,ae62}, barrier heights and reaction 
energies (K9) \cite{ae62,k9}, bond lengths of small molecules (MGBL) \cite{mgbl19}, lattice constants (LC29), 
bulk moduli (BM29), and cohesive energies (CE29) of 29 solids \cite{mattsson08,ss}.}
\end{figure}

In Fig. \ref{scan_fig} we report
the mean absolute errors divided by the PBE-TCA one, as functions of $\alpha$, 
for the family of XC functionals build from the exchange of Eq. (\ref{aaa})
and the TCA correlation. Several properties are considered:
atomization energies of small molecules (AE6) \cite{ae6,ae62},
barrier heights and reaction energies (K9) \cite{ae62,k9},
bond lengths of small molecules (MGBL) \cite{mgbl19},
lattice constants (LC29), bulk moduli (BM29), and
cohesive energies (CE29) of 29 solids \cite{mattsson08,ss}.
The plot confirms the findings of previous investigations
about PBE-TCA and SOL-TCA. In addition, it shows
that the evolution with $\alpha$ is not monotonic for all properties.
In fact, for atomization and cohesive energies minima located
approximately at $\alpha=0.11$ and $\alpha=0.25$ are observed.
This fact, together with the opposite behavior of molecular and solid-state
properties, indicates that a ``best'' compromise may only be found
for some intermediate values of $\alpha$. 
We observe that the INT-TCA functional seems to be close
to such  ``best'' choice. Thus, we simply assume the INT-TCA functional 
as our guess for the ``best compromise''.

The results of Fig. \ref{scan_fig} are also confirmed 
when additional systems and properties are analyzed.
This is the case of Table \ref{tab_a1} where we report the MAEs of 
different tests of relevance for computational chemistry and 
solid-state physics as obtained using the PBE, PBEint and PBEsol
functionals as well as their TCA variants.
\begin{table*}
\begin{center}
\caption{\label{tab_a1}Mean absolute errors for different tests, as obtained from various exchange-correlation functionals. 
For each couple of functionals the difference between the value obtained using the TCA and the original PBE-like correlation
is reported in the columns denoted by $\Delta$. The logarithmic relative absolute error (LRAE; Eq. (\ref{eq_lrae}))
is also reported at the bottom of each group of tests as well as for the global set.}
\begin{ruledtabular}
\begin{tabular}{lrrrrrrrrr}
Test  &  PBE  &  PBE-TCA  &  ($\Delta$)  &  PBEint  &  INT-TCA  &  ($\Delta$)  &  PBEsol  &  SOL-TCA  &  ($\Delta$)  \\ 
\hline
\multicolumn{10}{c}{Main group thermochemistry (kcal/mol)}\\
Atomiz. energy (AE6) &  13.35  &  10.59  &  (-2.76)  &  23.60  &  9.37  &  (-14.22)  &  33.21  &  17.95  &  (-15.26)  \\ 
Atomiz. energy (W4)  &  10.72  &  7.82  &  (-2.90)  &  15.49  &  9.37  &  (-6.12)  &  21.35  &  12.68  &  (-8.67)  \\ 
Atomiz. energy (G2/97) &  14.76  &  9.89  &  (-4.87)  &  26.44  &  11.14  &  (-15.30)  &  37.69  &  19.05  &  (-18.63)  \\ 
Reaction energy (BH76RC) &  4.35  &  3.96  &  (-0.40)  &  5.38  &  5.14  &  (-0.24)  &  6.31  &  6.21  &  (-0.10)  \\ 
Reaction energy (OMRE) &  6.74  &  7.05  &  (0.32)  &  8.05  &  6.32  &  (-1.73)  &  11.85  &  9.53  &  (-2.32)  \\ 
Barrier heights (BH76)  &  9.77  &  8.72  &  (-1.05)  &  10.90  &  9.55  &  (-1.35)  &  12.17  &  10.71  &  (-1.45)  \\ 
Kinetics (K9)  &  7.47  &  6.45  &  (-1.01)  &  9.05  &  8.04  &  (-1.01)  &  10.55  &  9.59  &  (-0.96)  \\ 
LRAE  &  -4.1  &  -11.9  &  (-7.8)  &  9.6  &  -8.0  &  (-17.5)  &  20.9  &  6.8 & (-14.1) \\ 
  &    &    &    &    &    &    &    &    &    \\  
\multicolumn{10}{c}{Main group geometry (m\AA{} and cm$^{-1}$)}\\
H bond lengths (MGHBL9) &  11.45  &  4.77  &  (-6.67)  &  14.72  &  6.95  &  (-7.77)  &  14.53  &  6.99  &  (-7.54)  \\ 
non-H bond lengths (MGNHBL11) &  7.64  &  7.44  &  (-0.20)  &  7.18  &  6.89  &  (-0.29)  &  5.19  &  6.04  &  (0.85)  \\ 
vibrations (F38) &  58.38  &  36.63  &  (-21.75)  &  66.65  &  40.27  &  (-26.38)  &  67.54  &  39.54  &  (-28.00)  \\ 
LRAE  &  8.2  &  -11.6  &  (-19.8)  &  12.9  &  -5.9  &  (-18.8)  &  8.2  &  -8.0  & (-16.2) \\ 
  &    &    &    &    &    &    &    &    &    \\ 
\multicolumn{10}{c}{Transition metals (kcal/mol, m\AA, and kcal/(mol$\cdot$atom) for AUnAE)}\\
Atomiz. energy (TM10AE) &  13.02  &  11.04  &  (-1.98)  &  15.20  &  11.16  &  (-4.04)  &  18.34  &  12.66  &  (-5.68)  \\ 
Reaction energy (TMRE) &  3.73  &  3.20  &  (-0.53)  &  6.94  &  4.64  &  (-2.29)  &  9.89  &  7.59  &  (-2.30)  \\ 
Gold cluster AE (AUnAE) &  0.60  &  2.30  &  (1.70)  &  1.50  &  0.81  &  (-0.69)  &  3.63  &  1.61  &  (-2.02)  \\ 
Bond lengths (TMBL) &  13.51  &  10.43  &  (-3.08)  &  17.28  &  17.01  &  (-0.28)  &  22.42  &  24.23  &  (1.81)  \\ 
Gold clusters BL (AuBL6) &  56.46  &  31.42  &  (-25.03)  &  22.12  &  25.63  &  (3.51)  &  21.11  &  25.74  &  (4.63)  \\ 
LRAE  &  -8.8  &  -7.2  &  (1.6)  &  -0.1  &  -10.4  &  (-10.3)  &  14.1  &  4.0  & (-10.1) \\ 
  &    &    &    &    &    &    &    &    &    \\ 
\multicolumn{10}{c}{Non-covalent interactions (kcal/mol)}\\
Hydrogen bonds (HB6) &  0.38  &  0.46  &  (0.08)  &  0.50  &  0.62  &  (0.12)  &  1.65  &  0.67  &  (-0.98)  \\ 
Dipole-dipole (DI6)  &  0.38  &  0.46  &  (0.08)  &  0.44  &  0.57  &  (0.13)  &  0.96  &  0.47  &  (-0.49)  \\ 
Dihydrogen bonds (DHB23) &  0.98  &  0.68  &  (-0.29)  &  1.01  &  0.66  &  (-0.35)  &  1.76  &  1.08  &  (-0.68)  \\ 
LRAE  &  -23.0  &  -22.7  &  (0.3)  &  -16.4  &  -15.7  &  (0.7)  &  20.2  &  -10.1 & (-30.4) \\ 
  &    &    &    &    &    &    &    &    &    \\ 
\multicolumn{10}{c}{Other molecular properties (kcal/mol)}\\
Multireference AE (W4-MR) &  21.80  &  14.90  &  (-6.90)  &  28.98  &  21.68  &  (-7.29)  &  35.50  &  27.74  &  (-7.77)  \\ 
Difficult cases (DC9/12)  &  40.75  &  29.78  &  (-10.97)  &  63.48  &  35.00  &  (-28.47)  &  82.94  &  53.18  &  (-29.76)  \\ 
Small interfaces (SI12) &  3.72  &  6.45  &  (2.72)  &  2.69  &  3.38  &  (0.69)  &  3.79  &  2.76  &  (-1.03)  \\ 
LRAE  &  -4.1  &  -6.2  &  (-2.1)  &  1.7  &  -7.8  &  (-9.6)  &  13.5  &  -1.1  & (-14.6) \\  
  &    &    &    &    &    &    &    &    &    \\ 
\multicolumn{10}{c}{Solid-state (m\AA, GPa, and eV)}\\
Lattice constants (LC29) & 53.00 & 63.29 & (10.28) & 36.05 & 40.72 & (4.67) & 30.99 & 29.84 & (-1.14) \\
Bulk moduli (BM29) & 12.43 & 13.87 & (1.43) & 8.87 & 9.26 & (0.39) & 8.14 & 8.86 & (0.71) \\
Cohesive energies (CE29) & 0.13	& 0.32 & (0.19) & 0.20 & 0.20 & (0.00) & 0.34 & 0.21 & (-0.13) \\
LRAE & -3.0 & 13.9 & (16.9) & -8.1 & -5.6 & (2.5) & -3.5 & -9.9 & (-6.4) \\
  &    &    &    &    &    &    &    &    &    \\ 
\multicolumn{10}{c}{Overall statistics}\\
Chemistry LRAE & -6.8  &  -12.7  &  (-5.9)  &  3.2  &  -10.3  &  (-13.5)  &  18.0  &  0.5  &  (-17.5)  \\ 
Solid-state LRAE & -2.7 & 14.2 & (16.9) & -7.8 & -5.3 & (2.5) & -3.2 & -9.5 & (-6.4) \\
Average LRAE & -4.9 & 0.6 & (5.5) & -2.5 & -8.0 & (-5.5) & 7.3 & -4.7 & (-12.0) \\
\end{tabular}
\end{ruledtabular}
\end{center}
\end{table*}
The data show that indeed PBE-TCA is the best functional for properties related to
computational chemistry (LRAE of -12.7), being especially good for thermochemistry,
structural properties, and non-covalent interactions. On the contrary, PBE-TCA is
the worst functional for solid-state properties.
An opposite behavior is found for SOL-TCA, which displays the best performance for
solid-state tests (LRAE=-9.5), outperforming also the original PBEsol; however, it
shows rather poor results for chemistry-related problems (LRAE=0.5). 
Note that anyway
SOL-TCA performs in this case definitely better than PBEsol and 
even better than PBEint (see below). 
Finally, INT-TCA shows a performance close to PBE-TCA
for computational chemistry and close to SOL-TCA for solid-state, 
being on average the best functional. 

A further interesting feature to observe is the comparison between
the original functionals using PBE-like correlation and their corresponding
variant using the TCA correlation. In this way, it is actually possible to 
understand whether the use of TCA may bring some advantages for a given
choice of exchange and a given property.
Such a comparison is summarized in Table \ref{tab_a1} by the quantity
denoted $\Delta$ which corresponds, for each line, to the difference
between the -TCA value and the PBE-like one.
Thus, negative values of $\Delta$ indicate that the use of TCA improves
the results, while positive values denote the opposite.
The values of $\Delta$ in the table show that the use of TCA correlation
systematically improves the description of chemistry-related properties,
with only few exceptions. This behavior is the same for all the 
considered exchange functionals. On the contrary, in general
the description of solid-state properties is slightly
worsened when TCA correlation is used. This is
not true, however, for the PBEsol exchange, since in this case 
both lattice constants and bulk moduli are left basically unchanged (within
the numerical precision) while the cohesive energies are improved.
Summarizing, we can say that, when TCA correlation is used 
in association with PBEint or PBEsol exchange, the improvement
for computational chemistry tests is bigger than the worsening for solid-state
tests (if any). Hence, these TCA-based functionals perform
better than their original counterparts. Oppositely, this is
not the case when PBE exchange is used. In such a case, the PBE-TCA
large improvement in chemistry-related
properties is accompanied by a significant worsening in solid-state
tests.

To complete our investigation on the compatibility of the TCA
correlation with GGA exchange functionals, we consider
its use with functionals not belonging to the family defined in Eq. (\ref{aaa}).
Of course, there exists a huge number of GGA exchange functionals and
any choice will be invariably arbitrary. Nevertheless, to take an easy option,
we have selected a few functionals that have originally been developed 
as exchange-only functionals, namely B88 \cite{b88}, 
OPTX \cite{optx}, Wu-Cohen \cite{wuco}, and PBEsol$_b$ \cite{sll}. 
These have been used to
form the corresponding XC functionals B-TCA, O-TCA, WC-TCA, and SOL$_b$-TCA.
The performance of these functionals for our suite of tests 
is reported in Table \ref{tab_a2}.
\begin{table*}
\begin{center}
\caption{\label{tab_a2}Mean absolute errors for different tests, as obtained from various exchange-correlation functionals using the TCA correlation. The logarithmic relative absolute error (LRAE; Eq. (\ref{eq_lrae})) is also reported at the bottom of each group of tests as well as for the global set. The label n.c. denotes non-converged calculations.}
\begin{ruledtabular}
\begin{tabular}{lrrrr}
Test  &  B-TCA  &  O-TCA  &  WC-TCA  &  SOL$_b$-TCA  \\ 
\hline
\multicolumn{5}{c}{Main group thermochemistry (kcal/mol)}\\
Atomiz. energy (AE6)  &  17.53  &  16.08  &  12.36  &  16.50  \\ 
Atomiz. energy (W4) &  9.00  &  9.82  &  10.45  &  13.43  \\ 
Atomiz. energy (G2/97)  &  16.72  &  14.07  &  13.42  &  16.67  \\ 
Reaction energy (BH76RC)  &  3.37  &  3.25  &  5.47  &  5.97  \\ 
Reaction energy (OMRE) &  9.08  &  6.36  &  6.64  &  8.44  \\ 
Barrier heights (BH76)  &  7.07  &  4.77  &  10.03  &  10.34  \\ 
Kinetics (K9)  &  4.79  &  3.29  &  8.59  &  9.19  \\ 
LRAE  &  -7.3  &  -15.5  &  -3.0  &  4.3  \\
  &    &    &    &    \\  
\multicolumn{5}{c}{Main group geometry (m\AA{} and cm$^{-1}$)}\\
H bond lengths (MGHBL9) &  4.59  &  3.05  &  13.30  &  6.88  \\ 
non-H bond lengths (MGNHBL11)  &  7.49  &  6.98  &  7.67  &  6.11  \\ 
vibrations (F38) &  35.80  &  34.70  &  66.00  &  38.96  \\ 
LRAE  &  -12.3  &  -19.8  &  12.2  &  -8.2  \\ 
  &    &    &    &    \\ 
\multicolumn{5}{c}{Transition metals (kcal/mol, m\AA, and kcal/(mol$\cdot$atom) for AUnAE)}\\
Atomiz. energy (TM10AE) &  10.93  &  10.17  &  12.16  &  12.22  \\ 
Reaction energy (TMRE) &  3.16  &  4.79  &  5.49  &  6.87  \\ 
Gold cluster AE (AUnAE) &  3.98  &  7.89  &  1.13  &  1.12  \\ 
Bond lengths (TMBL)  &  11.22  &  16.85  &  12.76  &  22.53  \\ 
Gold clusters BL (AuBL6) &  29.76  &  32.24  &  24.51  &  27.49  \\ 
LRAE  &  -2.5  &  10.7  &  -8.3  &  -0.4  \\ 
  &    &    &    &    \\  
\multicolumn{5}{c}{Non-covalent interactions (kcal/mol)}\\
Hydrogen bonds (HB6) &  2.15  &  4.40  &  0.65  &  0.52  \\ 
Dipole-dipole (DI6) &  1.84  &  3.01  &  0.50  &  0.56  \\ 
Dihydrogen bonds (DHB23) &  0.82  &  2.01  &  1.22  &  0.88  \\ 
LRAE  &  22.4  &  52.9  &  -8.0  &  -14.3  \\ 
  &    &    &    &    \\  
\multicolumn{5}{c}{Other molecular properties (kcal/mol)}\\
Multireference AE (W4-MR) &  10.66  &  8.95  &  23.69  &  26.30  \\ 
Difficult cases (DC9/12)  &  40.94  &  44.62  &  41.01  &  48.52  \\ 
Small interfaces (SI12) &  9.03  &  11.93  &  2.92  &  2.75  \\ 
LRAE  &  -1.6  &  1.2  &  -6.3  &  -3.2  \\ 
  &    &    &    &    \\ 
\multicolumn{5}{c}{Solid-state (m\AA, GPa, and eV)}\\
Lattice constants (LC29) & 70.92 & 73.35 & 34.46 & 34.03 \\
Bulk moduli (BM29) & 14.41 & 10.66 & 8.28 & 9.70 \\
Cohesive energies (CE29) & n.c. & 0.53 & 0.16 & 0.22 \\
LRAE & - & 19.2	& -12.9 & -5.9 \\ 
  &    &    &    &    \\ 
\multicolumn{5}{c}{Overall statistics}\\
Chemistry LRAE & -2.0  &  2.5  &  -3.6  &  -2.6  \\ 
Solid-state LRAE & - & 19.5	& -12.6 & -5.1 \\ 
Average LRAE & - & 10.9 & -8.3 & -4.2 \\
\end{tabular}
\end{ruledtabular}
\end{center}
\end{table*}
Inspection of the table shows that both B-TCA and O-TCA perform
in general rather poorly with the notable exception of structural
properties where they yield the best results among all the functionals
considered in this paper.
On the other hand, WC-TCA displays a very good performance for most tests (LRAE=-8).
This value is similar to the one of INT-TCA.
This similarity may be rationalized considering that both the 
exchange functionals perform an interpolation between the
slowly- and the rapidly-varying density regime. 
Finally, we note that SOL$_b$-TCA gives a global
LRAE worse than INT-TCA or WC-TCA but  
rather close to SOL-TCA.
Nevertheless, unlike the latter,  SOL$_b$-TCA shows a very balanced performance
among all tests, with no evident failures.
Thus, INT-TCA, WC-TCA, and SOL$_b$-TCA are the only XC functionals
displaying a negative LRAE both for chemistry and solid-state tests.

\section{Example applications}
To conclude our work, we discuss two
applications where TCA correlation can be 
fruitfully employed. To do this, we
apply the INT-TCA, WC-TCA, and SOL$_b$-TCA functionals
(the ones displaying negative LRAE values for both
chemistry and solid-state) to two 
challenging tests for GGA functionals: the calculation of
chemisorption energies $E_{ads}$ of CO on Pt(111) surfaces vs 
the corresponding surface energies $E_\sigma$ \cite{co1,co2}
and the computation of the interaction energy between a methylthiolate molecule
and a copper cluster Cu$_{17}$ \cite{pbeint}. These problems
require good accuracy for both
chemical and solid-state properties and are usually poorly
described by GGA functionals, which turn out to be too
``specialized'' to yield the
required balanced description of the problem.

\begin{figure}
\includegraphics[width=\columnwidth]{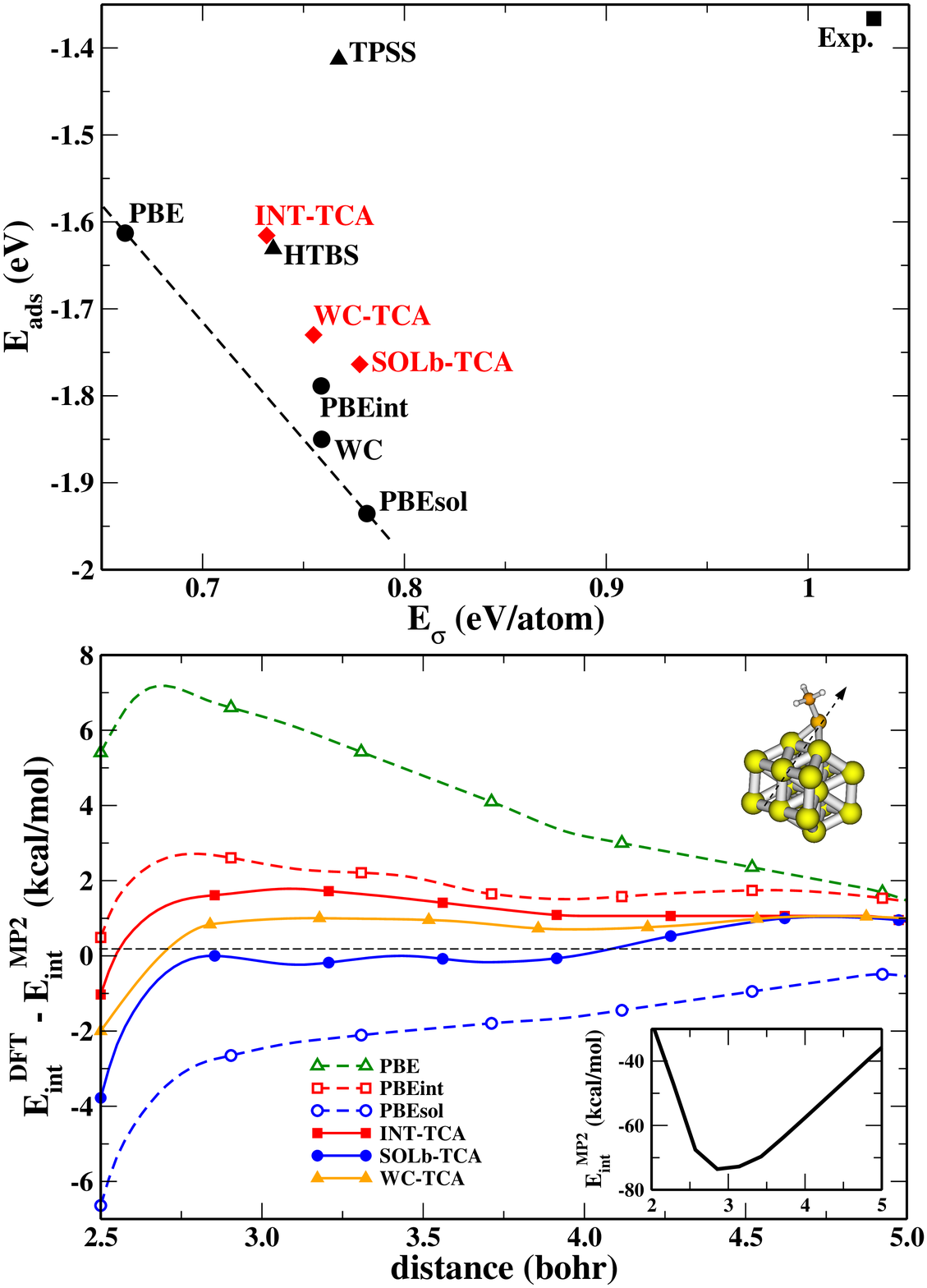}
\caption{\label{fig_final} Upper panel: chemisorption energy of CO on Pt(111)  as a function of the 
corresponding surface energy as obtained from several functionals. The dashed line indicates the usual behavior 
of most GGA functionals. Data different from INT-TCA, WC-TCA, and SOL$_b$-TCA ones are taken from Refs. 
\cite{co1,co2}. Lower panel: interaction energy errors for a methylthiolate molecule on a Cu$_{17}$ cluster as 
resulting from different functionals. Reference data are MP2 results \cite{pbeint}.}
\end{figure}

The results of our calculations are shown in Fig. \ref{fig_final}.
The upper panel reports the chemisorption energy of CO on Pt(111) versus  the
corresponding surface energy as obtained from several functionals. 
As shown in previous works \cite{co1,co2}, the results
of most GGA functionals tend to fall on a straight line, roughly defined by 
the PBE and PBEsol points (dashed line in the figure). This happens because,
in most cases, at the GGA level, improvements for the adsorption energy
come at the expense of the accuracy in computing the surface energies or
vice-versa.
Only few GGA functionals, specifically designed to balance the behavior
in different situations (e.g. PBEint \cite{pbeint} or HTBS \cite{co2}), 
can provide some improvement over the usual behavior. 
Otherwise, higher level functionals need to be considered \cite{co1}
(e.g the meta-GGA TPSS \cite{tpss} or non-local functionals).

Inspection of the figure shows that also INT-TCA, WC-TCA, and
SOL$_b$-TCA can break the usual trend, providing a
small, but significant, improvement with respect to most GGA functionals.
In particular, SOL$_b$-TCA yields a surface energy as accurate as
PBEsol but a much better adsorption 
energy. Thus, it finally performs slightly better than PBEint.
A similar behavior is observed for WC-TCA and INT-TCA.
For these functionals we can also note that the use of
the TCA correlation in place of the original ones leads to only a
modest worsening of the surface energy but a significant improvement
of the adsorption energy.

In the lower panel of Fig. \ref{fig_final} we report the
interaction
energy of a small molecule (SCH$_3$) with a metallic cluster
(Cu$_{17}$).
This rather simple model system constitutes 
a difficult test for GGA functionals, since it requires
a good description of both molecular and slowly-varying systems
\cite{pbeint}.
The curves reported in Fig. \ref{fig_final}
show that the SOL$_b$-TCA, WC-TCA, and PBEint-TCA
functionals perform well for this problem, as they give
a well balanced treatment of all density regimes. Thus,
they can finally outperform more specialized functionals such as PBE
and PBEsol.

\section{Conclusions}
We have performed a thorough investigation of the TCA correlation functional.
We have shown that, even if it does not recover the exact LDA limit, 
the TCA correlation performs well for solid-state problems.
Consequently, the TCA functional shows a broad applicability, being accurate for a variety of 
systems ranging from atoms to jellium surfaces. 
To check the compatibility of the TCA correlation
with GGA exchange functionals, we have studied its performance in conjunction
with a family of PBE-like exchange functionals as well as with
some other popular GGA exchange functionals.
We have found that the combination of the TCA correlation
with non-empirical exchange functionals gives
a significant improvement in the description of molecular
systems, but displays some limitations in the case of bulk solids.
Nevertheless, a good accuracy and a large applicability
can be achieved when the exchange part is well calibrated to
balance slowly- and rapidly-varying effects appropriately.

In conclusion, the present study indicates that 
TCA is competitive with other
state-of-the-art GGA correlation functionals,
showing its usefulness for electronic structure studies.
Furthermore, our assessment suggests that the WC-TCA, INT-TCA, and SOL$_b$-TCA XC
functionals are good tools for practical applications at the GGA level of theory.
These functionals have shown good average accuracy and  
wide applicability, being above-the-average for
both chemical and solid-state tests. 
Thus, they appear to be suitable computational tools to treat complex problems,
where different density regimes are involved,
e.g. etherogeneous catalysis.

\acknowledgements
We thank TURBOMOLE GmbH for providing the TURBOMOLE program package. 
E. Fabiano acknowledges the partial funding of this work 
from a CentraleSup\'elec visiting professorship.

\end{document}